\documentclass[twocolumn,aps,pra,amsmath,amssymb,showpacs]{revtex4}
\usepackage[latin9]{inputenc}
\usepackage{amsmath}
\usepackage{amssymb}
\usepackage{graphicx}
\usepackage{esint}

\makeatletter
\@ifundefined{textcolor}{}
{%
 \definecolor{BLACK}{gray}{0}
 \definecolor{WHITE}{gray}{1}
 \definecolor{RED}{rgb}{1,0,0}
 \definecolor{GREEN}{rgb}{0,1,0}
 \definecolor{BLUE}{rgb}{0,0,1}
 \definecolor{CYAN}{cmyk}{1,0,0,0}
 \definecolor{MAGENTA}{cmyk}{0,1,0,0}
 \definecolor{YELLOW}{cmyk}{0,0,1,0}
}

\usepackage{epsfig}\usepackage{amsfonts}

\makeatother

\begin{document}

\title{Phonon induced spin squeezing based on geometric phase }

\author{Yan-Lei Zhang, $^{1,2}$ }

\author{Chang-Ling Zou, $^{1,2,3}$ }

\email{clzou321@ustc.edu.cn}

\author{Xu-Bo Zou, $^{1,2}$ }

\email{xbz@ustc.edu.cn}

\author{Liang Jiang, $^{3}$}

\author{Guang-Can Guo $^{1,2}$ }

\affiliation{$^{1}$ Key Laboratory of Quantum Information, University of Science
and Technology of China, Hefei, 230026, People's Republic of China; }

\affiliation{$^{2}$ Synergetic Innovation Center of Quantum Information \& Quantum
Physics, University of Science and Technology of China, Hefei, Anhui
230026, China}

\affiliation{$^{3}$ Department of Applied Physics, Yale University, New Haven,
CT 06511, USA}

\date{\today}
\begin{abstract}
A scheme to achieve spin squeezing using a geometric phase induced
by a single mechanical mode is proposed. The analytical and numerical
results show that the ultimate degree of spin squeezing depends on
the parameter $\frac{n_{th}+1/2}{Q\sqrt{N}}$, which is the ratio
between the thermal excitation, the quality factor and square root
of ensemble size. The undesired coupling between the spin ensemble
and the bath can be efficiently suppressed by Bang-Bang control pulses.
With high quality factor, the ultimate limit of the ideal one-axis
twisting spin squeezing can be obtained for an NV ensemble in diamond.
\end{abstract}


\maketitle
\emph{Introduction.}- The NV centers in diamond are amongst the most
promising implementations of quantum bits for quantum information
processing \cite{Childress2013} and nanoscale sensors \cite{Rondin2014},
which is because their ground state spin triplet posses ultra-long
coherent time at room temperature \cite{Balasubramanian2009} and
can be readout via optical fluorescence. Significant progresses have
been achieved in recent experiments to couple the NV electronic spins
to nuclear spins \cite{Jelezko2004,Childress2006} and mechanical
resonators \cite{Kolkowitz2012,Ovartchaiyapong2014}. The nanoscale
magnetometry \cite{Maze2008,Pham2011}, thermometer \cite{Kucsko}
and electric field detection \cite{Dolde} have been demonstrated
by single NV or an ensemble.

It's well known that the quantum states can boost the precision measurement
beyond the standard quantum limit \cite{Giovannetti}. Among them,
the spin squeezed states (SSS) \cite{Kitagawa1993,Itano1993,Wineland1994,Ma2011}
have attracted a lot of interest and applied to spin or atom ensembles
for atomic clocks and gravitational wave interferometers. There are
many proposals and experiments to realize the spin squeezing in atom
ensembles, such as atom-atom collisions \cite{Riedel2010}, quantum
non-demolition (QND) measurement \cite{Chaudhury2007,Inoue2013} and
cavity squeezing \cite{Ueda1996,Takeuchi2005,Schleier2010,Leroux2010,Torre2013,Zhang2014}.
Very recently, spin squeezing of an NV ensemble by Tavis-Cummings
type interaction between phonon and spins \cite{Bennett2013} has
been proposed for quantum enhanced magnetometry.

In this paper, we propose a new approach for the realization of spin
squeezing by phonon induced geometric phase, using an ensemble of
NV centers dispersively coupled to a mechanical resonator. It's shown
that the ultimate degree of spin squeezing by one-axis twisting can
be realized, for reasonable ratio between the thermal excitation and
the quality factor of mechanical oscillators. Furthermore, the effect
of the coupling between NV centers and environment is studied, which
leads to dephasing and degrades the spin squeezing effect. By introducing
Bang-Bang pulses sequence, the decoherence is effectively suppressed
and significant spin squeezing can be achieved for the NV ensemble.

\emph{Model.-} The negatively charged NV center (NV$^{-}$) in diamond
is well-studied, whose Hamiltonian reads $H_{NV}=(D+d^{\parallel}\epsilon_{z})S_{z}^{2}+\mu_{b}g_{e}\overrightarrow{S}\cdot\overrightarrow{B}$
\cite{Dolde,Fang}, where $D\approx2.87$ GHz is zero-field splitting,
$d^{\parallel}$ and $\epsilon_{z}$ are axial ground-state electric
dipole moment and electric field (strain field), respectively. With
appropriate bias field $B_{z}$, the two microwave transitions $\left|0\right\rangle \leftrightarrow\left|\pm1\right\rangle $
can be addressed separately in experiment, and we focus on the $\left|0\right\rangle \leftrightarrow\left|-1\right\rangle $,
which transition can be treated as a spin-$\frac{1}{2}$ system in
the following. Putting the NV$^{-}$ spin ensemble in a gradient magnetic
field $\frac{\partial B_{z}}{\partial u}\neq0$, then the displacement
of diamond or nanomagnet $\delta u$ will shift the transition frequency
by $\Delta\omega_{NV}=\mu_{b}g_{e}\sigma_{z}\frac{\partial B_{z}}{\partial u}\delta u$
\cite{Rugar2004}. Alternatively, the strain field of a diamond nanomechanical
oscillator can induce an electric field inside the crystal and give
rise to a similar phonon-spin interaction \cite{Ovartchaiyapong2014,Teissier2014}.
Both approaches to couple the spin with nanomechanical oscillator
have been demonstrated in experiments recently \cite{Ovartchaiyapong2014,Kolkowitz2012,Rugar2004,Teissier2014,Ganzhorn2013,Tian2014}.
The simplified Hamiltonian of an ensemble of $2N$ spins coupled to
a mechanical resonator is
\begin{equation}
H=\omega_{a}a^{\dagger}a+gJ_{z}\left(a+a^{\dagger}\right),\label{eq:1}
\end{equation}
where $a$ and $a^{\dagger}$ are annihilation and creation operators
of phonon, $\omega_{a}$ is the frequency of the mechanical resonator,
$J_{z}=\frac{1}{2}\sum_{j=1}^{2N}\sigma_{zj}$ is the collective spin
operator, and $g$ is the single phonon coupling strength. Along with
the progresses in the nanofabrication of diamond material, various
diamond nanomechanical resonators have been realized in experiment,
with frequency ranging from 1 kHz to 1 GHz, and the quality factor
$Q$ ranging from 100 to around $10^{6}$ \cite{Ovartchaiyapong2014,Teissier2014,Burek2013,Zalalutdinov2011,Tao2014}.
Thus, we study the spin squeezing induced by the mechanical resonator
with frequency $\omega_{a}/2\pi=1$ MHz and coupling strength $g/2\pi=1$
kHz \cite{Rabl} in this work.

\begin{figure}
\includegraphics[width=1\columnwidth]{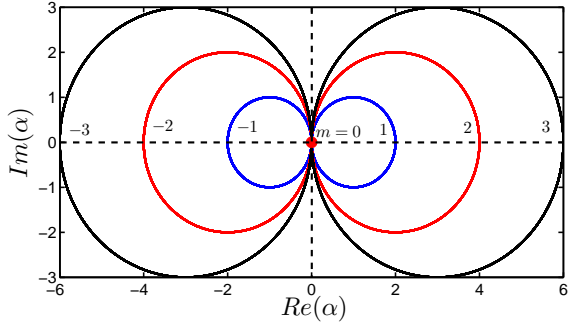}

\protect\caption{Trajectories on phase space of a coherent wave packet $\left\langle a\right\rangle $=$\mathrm{Re}(\alpha)+i\mathrm{Im}(\alpha)$
for spin state $\left|m\right\rangle $ with $m=0,~\pm1,~\pm2,~\pm3$.
Here we set $g/\omega_{a}$ as a unit.}
\end{figure}

The Hamiltonian preserves $J_{z}$ of the spin ensemble. When we integrate
the Schrodinger equation for collective spin states, the phonon will
introduces a spin dependent geometric phase shift \cite{Puri2012}.
It is convenient to study the mechanical resonator by the coherent
state $\left|\alpha\right\rangle $, and we can write \textbf{$\left|\alpha\right\rangle =\left|\mathrm{Re}(\alpha)+i\mathrm{Im}(\alpha)\right\rangle $}
with $\alpha(t)=\frac{-gJ_{z}}{\omega_{a}}\left[1-e^{-i\omega_{a}t}\right]$.
The coherent state behaves somehow like classical particles in phase
space. Its center, given by $\mathrm{Re}(\alpha)$ and $\mathrm{Im}(\alpha)$,
follows a classical trajectory, while the width of these wave packets
remains fixed, which is given by the uncertainty of the $\mathrm{Re}(\alpha)$
and $\mathrm{Im}(\alpha)$. In Fig. 1 we plot the usual phase-space
trajectories for $\left\langle a\right\rangle =\mathrm{Re}(\alpha)+i\mathrm{Im}(\alpha)$,
and we have used the eigenstates $\left|m\right\rangle $ of spin
operator $J_{z}$ as the initial states of the collective spin. We
plot phase-space trajectories only with $m=0,~\pm1,~\pm2,~\pm3$ for
simple explanation, which clearly show that the coherent wave packet
is restored to its original state after a fixed time $T=2\pi/\omega_{a}$
or integer times of $T$. For different $\left|m\right|$, there are
different radius circles in the phase-space trajectories, and it is
the central symmetry for the opposite $m$.

The geometric phase, as the enclosed circle area of the trajectory
in phase space, is insensitive to the initial state \cite{Garcia2005}.
Thus, phonon induced geometry phase is robust against the imperfection
of initial phonon state preparation, and we assume $\alpha\left(0\right)=0$
for simplicity. However, the decay and thermal noise of phonon during
the spin-phonon interaction will influence the geometry phase accumulation.
In this case, the system dynamics follows the Master equation
\begin{equation}
\frac{d\rho}{dt}=-i\left[H,\ \rho\right]+\frac{\gamma}{2}\left(n_{th}+1\right)\mathcal{L}(a)\rho+\frac{\gamma}{2}n_{th}\mathcal{L}(a^{\dagger})\rho.
\end{equation}
Here $\gamma=\omega_{a}/Q$ describes the decay rate of the mechanical
mode, $n_{th}$ is the mean phonon number of the mechanical thermal
noise and $\mathcal{L}(o)\rho=2o\rho o^{\dagger}-o^{\dagger}o\rho-\rho o^{\dagger}o$
is the Lindblad superoperator for given jump operator $o$. The reduced
density matrix of the collective spin can be written as
\begin{equation}
\rho_{spin}=\sum_{m,n}\rho_{m,n}(0)e^{\phi_{m,n}(t)}\left|m\right\rangle \left\langle n\right|.\label{eq:density2-1}
\end{equation}
 $\phi_{m,n}(t)$ is the phase difference between these spin states.
The phase can be solved as
\begin{eqnarray}
 &  & \phi_{m,n}(t)\nonumber \\
 & = & -\left(n_{th}+\frac{1}{2}\right)(n-m)^{2}\left\{ \gamma\int_{0}^{t}\mid\alpha(\tau)\mid^{2}d\tau+\mid\alpha(t)\mid^{2}\right\} \nonumber \\
 &  & +ig(n^{2}-m^{2})\mathrm{Re}\int_{0}^{t}\alpha(\tau)d\tau.\label{eq:phase}
\end{eqnarray}
Here, the amplitude of mechanical resonator is $\alpha(t)=\frac{-ig}{\gamma/2+i\omega_{a}}\left[1-e^{-\left(\gamma/2+i\omega_{a}\right)t}\right]$
\cite{Garcia2005}. The finite $\gamma$ of the mechanical resonator
introduces decoherence and leads to the first term of the above equation,
the second term is corresponding to the interaction $J_{z}^{2}$ inducing
spin squeezing. Molmer and Sorensen proposed an approach for ion-trap
to realize the spin squeezing, which is insensitive to the initial
thermal phonon states \cite{IonTrap}. Compare to the Molmer-Sorensen
scheme that two laser pumping and Lamb-Dicke approximation are required
\cite{IonTrap}, our approach utilize stable spin-phonon interaction
and there is no approximation in our model.

\emph{Spin squeezing.-} The spin squeezing is evaluated by squeezing
parameter \cite{Kitagawa1993,Ma2011}
\begin{equation}
\xi_{N}^{2}=\frac{min\left(\Delta J_{\vec{n}_{\perp}}^{2}\right)}{N/2},
\end{equation}
where $\Delta J_{\vec{n}_{\perp}}^{2}$ is the variance of spin operators
along direction perpendicular to the mean-spin direction $\vec{n_{0}}=\frac{\vec{J}}{|\langle\vec{J}\rangle|}$,
which is determined by the expectation values $\left\langle J_{\alpha}\right\rangle $,
with $\alpha\in\left\{ x,y,z\right\} $. For an atomic system initialized
in a coherent spin state (CSS) \cite{Arecchi1972} along the $x$
axis, satisfying $J_{x}\left|\psi\left(0\right)\right\rangle _{\mathrm{atom}}=N\left|\psi\left(0\right)\right\rangle _{\mathrm{atom}}$,
we have $\rho_{m,n}\left(0\right)=2^{-2N}\sqrt{\frac{(2N)!}{(N-m)!(N+m)!}\frac{(2N)!}{(N-n)!(N+n)!}}$
and $\Delta J_{\vec{n}_{\perp}}^{2}=N/2$. Thus, for squeezed spin
states we have $\xi_{N}^{2}<1$.

First of all, we studied the spin squeezing by Eq. (3) without thermal
noise. The squeezing parameters $\xi_{N}^{2}$ as a function of the
time (dimensional number $gt$) for various quality factor $Q$ are
plotted in Fig. 2(a). As expected, the effect of phonon induced geometry
phase leads to the twisting and squeezing of CSS, thus the $\xi_{N}^{2}$
decreasing with time. After a certain optimal $t$, the $\xi_{N}^{2}$
increases, due to the over twisting by the geometry phase, and high
order effect arises. It is shown that the minimal value of spin squeezing
parameter decreases with higher mechanical quality factor $Q$. When
the quality factor $Q=1000$ (the black solid line), the almost perfect
spin squeezing for the ideal one-axis twisting can be achieved. Including
the mechanical thermal noise, squeezing parameters $\xi_{N}^{2}$
as functions of the time with the quality factor $Q=1000$ are plotted
in Fig. 2(b). It is natural that the spin squeezing becomes worse
with the increasing of the thermal noise $n_{th}$. We also studied
the suppression of the influence of thermal noise by improving the
quality factor $Q$. As shown in Fig. 2(c), the optimal spin squeezing
(the minimum value of the $\xi_{N}^{2}(t)$) is plotted against the
$Q$ for $n_{th}=100$. The $\xi_{N}^{2}$ reduces with $Q$ and approaches
the limit for ideal one-axis twisting spin squeezing (red dashed line).
\begin{figure}
\includegraphics[width=1\columnwidth]{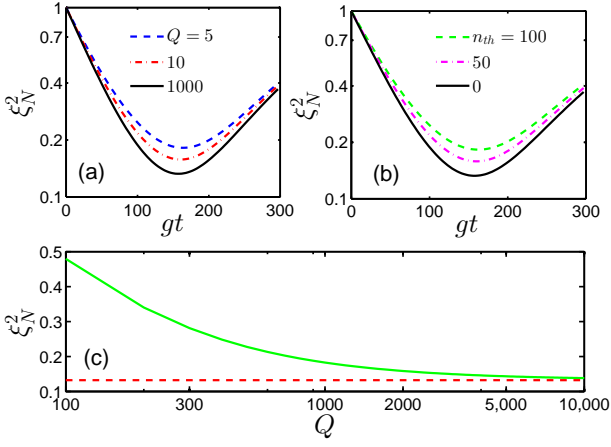}

\protect\caption{(Color online) (a) The squeezing parameter $\xi_{N}^{2}$ as a function
of the time for $n_{th}=0$ and various $Q=5$ , $10$ , $1000$ (top
to bottom). (b) The squeezing parameter $\xi_{N}^{2}$ as a function
of the time for $Q=1000$ and various $n_{th}=100$ , $50$ , $0$
(top to bottom). (c) The green solid line is optimal squeezing parameter
$\xi_{N}^{2}$ versus the quality factor $Q$ for $n_{th}=100$, and
the red dashed line is the result for ideal one-axis twisting spin
squeezing. $N=10$ for all simulations.}
\end{figure}

To understand these results, we simplified the spin state state dependent
geometric phase
\begin{equation}
\phi_{m,n}(t)=\frac{i|g|^{2}\omega_{a}t}{\left(\gamma/2\right)^{2}+\omega_{a}^{2}}\left[\left(m^{2}-n^{2}\right)+i\mu\left(m-n\right)^{2}\right],\label{eq:steady}
\end{equation}
under the approximation $t\gg\gamma^{-1}$, which means $\alpha(t)=\frac{-ig}{\gamma/2+i\omega_{a}}$
and the transient evolution of the mechanical resonator is neglected.
Here, the dimensionless factor $\mu=\frac{n_{th}+1/2}{Q}$. The first
term accounts for the coefficient proportional to the time $t$, and
the two terms within the bracket corresponds to spin squeezing and
decoherence, respectively. Then, we can obtain the degree of spin
squeezing for the initial state CSS along the $x$ axis
\begin{equation}
\xi_{N}^{2}=1+\frac{2N-1}{4}\left(A-\sqrt{A^{2}+B^{2}}\right),
\end{equation}
where
\begin{align}
A & =1-\cos^{2N-2}\left(2Ct\right)e^{-4C\mu t},\nonumber \\
B & =-4\sin\left(Ct\right)\cos^{2N-2}\left(Ct\right)e^{-4C\mu t}.
\end{align}
Here, $Ct=\frac{g}{\omega_{a}}\frac{1}{1+1/4Q^{2}}\times gt$. The
analytical solution implies that the spin squeezing is mainly determined
by the two dimensionless parameters $Ct$ and $\mu$. For $Q\gg1$,
we have $\frac{1}{1+1/4Q^{2}}\approx1$. For $N\gg1$, we can apply
the approximation $\cos^{2N-2}\left(x\right)\approx e^{-(N-1)x^{2}}$
for $x\ll1$. So, the time required ($gt\approx160$ in Fig. 2) to
achieve the optimal spin squeezing scales with $\frac{1}{\sqrt{N-1}}$.
From Eq. (7), we obtain the approximated upper bound of the optimal
spin squeezing $\xi_{N}^{2}\lesssim1-e^{-\frac{1}{2}-4\frac{\mu}{\sqrt{N}}}/(1-e^{-1-2\frac{\mu}{\sqrt{N}}})$,
which indicating that the ratio $\frac{\mu}{\sqrt{N}}=\frac{n_{th}+1/2}{Q\sqrt{N}}$
should be as small as possible. As long as $\frac{n_{th}+1/2}{Q\sqrt{N}}<10^{-3}$,
we can achieve squeezing almost as good as the best squeezing achievable
with ideal single axes twisting {[}Fig. 2(c){]}.

\emph{Bang-Bang control.-} During the preparation of optimal SSS,
there are inevitable couplings between the system and baths. For example,
the lattice vibrations and environment spins will induce dephasing
and destroy the spin squeezing. The dynamical decoupling technique
is well known for protecting coherence from environment \cite{Viola,Lidar,Du,Hanson,Jiang2011,RevDD,Tan2014},
and now we apply the Bang-Bang (BB) pulses \cite{Viola} to suppress
the decoherence. The sequence consists of $M$ pulses, which split
the total time interval $t$ into $M$ small intervals $t_{p}=\frac{p}{M}t$
with $p=1,2,...,M$. The pulses rotate the collective spin states
around $y$ axis, and we chose the pulse sequence to rotate $\pi$
and $-\pi$ alternately, which leads to $e^{i\pi J_{y}}\sigma_{zj}e^{-i\pi J_{y}}=-\sigma_{zj}$
and $e^{i\pi J_{y}}J_{z}^{2}e^{-i\pi J_{y}}=J_{z}^{2}$. Therefore,
the spin squeezing $J_{z}^{2}$ is conserved while the $\sigma_{z}$
is inverted by the BB pulses. Considering the $2N$ qubits which are
independently coupled to thermal baths, the Hamiltonian from the Eq.
(\ref{eq:1}) is changed to
\begin{eqnarray}
H^{\prime} & = & \omega_{a}a^{\dagger}a+g\varepsilon(t)J_{z}\left(a+a^{\dagger}\right)+\sum_{k}\omega_{k}b_{k}^{\dagger}b_{k}\nonumber \\
 &  & +\sum_{j=1}^{2N}\sum_{k}\frac{\varepsilon(t)\sigma_{zj}}{2}h_{kj}(b_{k}+b_{k}^{\dagger}).
\end{eqnarray}
Here, $b_{k}$ and $b_{k}^{\dagger}$ are the creation and annihilation
bosonic operators of the $k$-th bath mode, which coupling to the
$j$-th spin with coupling strength $h_{kj}$. The switch function
$\varepsilon(\tau)$ due to BB pulses is given by $\varepsilon(\tau)=\sum_{p=1}^{M}(-1)^{p}\theta(\tau-t_{p})\theta(t_{p+1}-\tau)$
with $\theta\left(t\right)$ is the Heaviside step function.
\begin{figure}
\includegraphics[width=1\columnwidth]{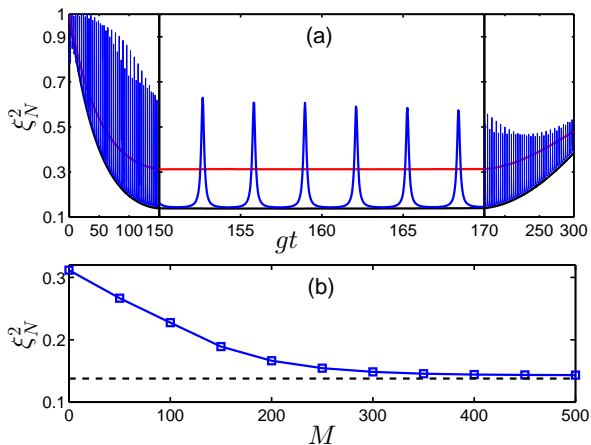}

\protect\caption{(Color online) (a) The squeezing parameter $\xi_{N}^{2}$ as a function
of the time $gt$. The parameters are $\eta=0$, $M=0$ (black line),
$\eta=4\times10^{-4}$, $M=0$ (red line), and $\eta=4\times10^{-4}$,
$M=500$ (blue line). (b) The blue solid line for the optimal squeezing
parameter $\xi_{N}^{2}$ versus the pulses $M$ for $\eta=4\times10^{-4}$,
and the black dashed line for the result without thermal baths. Other
parameters are $n_{th}=10$, $\lambda=4/g$, $Q=1000$ and $N=10$. }
\end{figure}

With the decoherence and BB, the geometry phase factor {[}Eq. (\ref{eq:phase}){]}
of spin states are solved as
\begin{eqnarray}
 &  & \phi_{m,n}^{\prime}(t)\nonumber \\
 & = & -\left(n_{th}+\frac{1}{2}\right)(n-m)^{2}\left\{ \gamma\int_{0}^{t}\mid\alpha'(\tau)\mid^{2}d\tau+\mid\alpha'(t)\mid^{2}\right\} \nonumber \\
 &  & +ig(n^{2}-m^{2})\mathrm{Re}\int_{0}^{t}\varepsilon(\tau)\alpha^{\prime}(\tau)d\tau-\kappa_{m,n}(t).
\end{eqnarray}
Here, $\alpha^{\prime}(t)=-ig\int_{0}^{t}\varepsilon(\tau)e^{-(\frac{\gamma}{2}+i\omega_{a})(t-\tau)}d\tau$
and $\kappa_{m,n}(t)$ is due to the decoherence. Assume that the
baths to each spin are Ohmic and have the same spectral density $\eta\omega e^{-\frac{\omega}{\omega_{c}}}$,
we have
\begin{equation}
\kappa_{m,n}(t)\leq(|n-m|+2)\int_{0}^{\infty}G\left(\omega\right)F_{M}(\omega,t)d\omega,
\end{equation}
where the modulation spectrum is $F_{M}(w,t)=\frac{\tan^{2}\left(\frac{\omega t}{2M+2}\right)\left(1+\left(-1\right)^{M}\cos\left(\omega t\right)\right)}{\omega^{2}}$,
the temperature-dependent interacting spectrum is $G(\omega)=\eta\omega e^{-\frac{\omega}{\omega_{c}}}\left(\frac{2}{e^{\lambda\omega}-1}+1\right)$
\cite{Tan2014}, in which $\eta$ is the coupling strength between
the system and the bath modes, $\omega_{c}$ is the cutoff frequency,
and $\lambda=1/\kappa_{B}T$ is the inverse temperature. In order
to simplify the calculation, we use the upper limit instead of the
$\kappa_{m,n}(t)$.

In Fig. 3(a), we\textbf{ }numerically calculated squeezing parameter
$\xi_{N}^{2}$ as a function of time for various $\eta$ and $M$.
Since the decoherence term $\kappa_{m,n}(t)$ is proportional to the
coupling strength between the system and the bath modes, we observe
the incremental of the optimal $\xi_{N}^{2}$ for increasing $\eta$
(black and red lines). The blue line shows the suppression of decoherence
by BB, and here we choose the sequence number $M=500$ and $\eta=4\times10^{-4}$,
which is contrast to the red line. There are periodic peaks with the
separation distance $\Delta gt=\pi$, and the peak values are obtained
when $gt=\left(n+1/2\right)\pi$, $n$ is integer. This phenomena
can be interpreted as following: the BB pulse period is $T_{M}=t/M$,
and the time period for the phonon state trajectories in the phase-space
{[}Fig. 1{]} is $T=2\pi/\omega_{a}$. For $M=500$ and $\omega_{a}/g=1000$,
we have $T_{M}/T=gt/\pi$. When $gt/\pi=n$ is integer, the geometric
phase is always cumulative, and the coherent spin squeezing effect
is not degraded by the BB pulse sequence. In contrast, when $gt/\pi=n+\frac{1}{2}$,
the geometric phase imprints alternating sign as function of $M$
, and then the spin squeezing is weakened. Compare the minimas of
$\xi_{N}^{2}$ with BB (blue line) to the results without BB (black
and red lines), the undesired effect of decoherence is effectively
suppressed by the dynamical decoupling. Fig. 3(c) shows the optimal
$\xi_{N}^{2}$ versus the pulse sequence length $M$.  With increasing
$M$, the optimal value of $\xi_{N}^{2}$ is improved and approaches
the red dashed line, which is the ideal result determined by the Eq
(\ref{eq:steady}) without thermal baths. When $M\geq400$, the influence
of thermal baths on spin squeezing can be almost eliminated, which
means $\kappa_{m,n}(t)\approx0$.

\emph{Conclusion.-} An approach to achieve spin squeezing by phonon
induced geometric phase is proposed. This scheme is feasible for experiments
on solid state spin ensemble coupled to a mechanical oscillator. With
reasonable parameters, the ultimate limit of the ideal one-axis twisting
spin squeezing can be achieved as long as the quality factor is sufficiently
high that $Q>\frac{n_{th}+1/2}{\sqrt{N}}\times10^{3}$. The decoherence
due to spin-bath coupling can be effectively suppressed by the Bang-Bang
pulses. This geometric-phase-based spin squeezing can be used to significantly
improve the sensitivity of magnetic sensing with nitrogen-vacancy
spin ensembles. Moreover, the technique can be generalized to spin
ensembles coupled to other high-Q Bosonic modes that prepare quantum
states by geometry phase.

{\em Acknowledgments.} This work was funded by National Basic Research
Program of China (Grant Nos. 2011CB921200 and 2011CBA00200) and National
Natural Science Foundation of China (Grant Nos. 11074244 and 11274295),
973project (2011cba00200). LJ acknowledges support from the Alfred
P. Sloan Foundation, the Packard Foundation, the AFOSR-MURI, the ARO,
and the DARPA Quiness program.

\end{document}